# SCALABLE LINK PREDICTION IN TWITTER USING SELF-CONFIGURED FRAMEWORK

Nur Nasuha Daud[1], Siti Hafizah Ab Hamid[1], Chempaka Seri[1], Muntadher Saadoon[1] and Nor Badrul Anuar[2]

[1]Department of Software Engineering, Faculty of Computer Science and Information Technology, University of Malaya, 50603, Kuala Lumpur, Malaysia
nasuha@um.edu.my, sitihafizah@um.edu.my, chempakaseri96@gmail.com, core93@yahoo.com

[2]Department of Computer System and Technology, Faculty of Computer Science and Information Technology, University of Malaya,50603, Kuala Lumpur, Malaysia
badrul@um.edu.my

*ABSTRACT*

*Link prediction analysis becomes vital to acquire a deeper understanding of events underlying social networks interactions and connections especially in current evolving and large-scale social networks. Traditional link prediction approaches underperformed for most large-scale social networks in terms of its scalability and efficiency. Spark is a distributed open-source framework that facilitate scalable link prediction efficiency in large-scale social networks. The framework provides numerous tunable properties for users to manually configure the parameters for the applications. However, manual configurations open to performance issue when the applications start scaling tremendously, which is hard to set up and expose to human errors. This paper introduced a novel Self-Configured Framework (SCF) to provide an autonomous feature in Spark that predicts and sets the best configuration instantly before the application execution using XGBoost classifier. SCF is evaluated on the Twitter social network using three link prediction applications: Graph Clustering (GC), Overlapping Community Detection (OCD), and Redundant Graph Clustering (RGD) to assess the impact of shifting data sizes on different applications in Twitter. The result demonstrates a 40% reduction in prediction time as well as a balanced resource consumption that makes full use of resources, especially for limited number and size of clusters.*

## 1. INTRODUCTION

Link prediction is essential for better understanding how individual interactions and connections evolve in social networking platforms. Statistics highlighted that current number of social network users are increasing linearly every year, where as of April 2020 there are almost 3.9 billion people were active internet users [1]. This evolution inspired many researchers over the past few years to explore new research areas of studies related to link prediction in large-scale social networks. Several efforts are available to address the issues of link prediction in large-scale social networks [2]. Hence, the framework in a distributed computing environment Spark was already successfully adopted in recent link prediction works for good prediction performance of large-scale social networks. Link prediction analysis resolved in less time with Spark by multiple computing resources given the in-memory computation, parallel job processing over master-slave architecture, and its scalability features. Spark also provides numerous properties to configure the computation process such as application properties.

Nonetheless, the difficulty of manually configuring the application properties for execution emerges when the performance of application is degraded and the resource utilization is imbalanced. Further, manual work is extremely difficult and time-consuming for users with less knowledge of how to use the framework. The configuration of properties is critical to the analysis that makes our processing system performs efficiently. Besides the need for efficient link prediction execution, the presence of automatically and correctly configured properties encouraged this research. Using the autonomic computing concept, we proposed a novel self-configured framework based on a trained XGBoost classifier to select the best configuration parameters suited to each submitted application in Spark. The proposed framework attempts to demonstrate performance and efficiency improvements in link prediction analysis in large-scale social networks while consuming resources efficiently and effectively.

The Self-Configured Framework (SCF) is further examined using three link prediction applications which include Graph Clustering (GC), Overlapping Community Detection (OCD), and Redundant Graph Detection (RGD) in Apache Spark. The evaluation results show that in terms of time performance, SCF is able to improve almost 40% of time performance in comparison with default configuration without SCF. Furthermore, we discovered that SCF contributes to balancing the resource utilization presented by the resource utilization rate. Accordingly, the next Section 2 provided an overview of related works in scalable link prediction. Then, in Section 3 we present the proposed framework with its implementation details. Section 4 presented the framework evaluation and highlighted the findings. Finally, Section 5 summarize the important key observations and conclude the proposed work.

## 2. RELATED WORKS

Link prediction research has evolved in recent years from single computing to distributed computing in an effort to provide scalable link prediction with today's vast data. CBRA [3] invented a novel MapReduce-based method for predicting future links proposed to achieve efficiency in large-scale networks. The parallel CBRA showed great efficiency compared with a traditional single computing link prediction. Although, its computation performance is limited due to the map-reduce procedure involving heavy I/O that consequently affects the prediction performance. Later, PCLP [4] proposed to support parallel link prediction using the Pregel model, a Bulk Synchronous Parallel (BSP) abstraction. BSP supports parallel computing adopted as a major technology for graph analytics at massive scale via Pregel and MapReduce [5]. Link prediction performs better when utilizing Pregel with a big data processing framework like Spark. DTLPLP [6] is the most recent work implementing scalable link prediction on Spark framework conducted using three real-world networks, Enron email, Collaboration Ca-Gr, and Facebook network. Majority of studies in scalable link prediction utilized the Pregel model in Spark framework to handle parallel and iterative processing on large data. The key benefit is that it makes algorithm implementation simple and provides scalability features for enormous datasets. Despite the success demonstrated, the parallel process entails a lot of message generations and transfers, which degrades system performance.

Additionally, there are prevailing works that proposed autonomic computing concept with self-configure feature in big data processing framework. Starfish [7] introduced self-tuning system on Hadoop based on user needs and system workloads for better performance using cost-based modeling and simulation. [8] used cron with python automated script to provide self-configure feature in Hadoop. The cron schedules its jobs automatically and scales the computation based on the load of the cluster for maximum efficiency of the cluster. Utilizing a cluster reconfiguration algorithm, [9] proved the algorithm is able to dynamically scale according to workload and conserve resource energy in cloud computation up to 54%. InSTechAH scales smart computing tasks on clusters automatically by using a workload prediction algorithm in a KVM-based cloud. There is still a scarcity of studies on how to correctly configure properties of the Spark framework. [10] presented an auto-tuning using a machine learning method, neural network model applied in Spark streaming applications to predict the increase or decrease of cluster configuration. The authors in [11] presented a simulation-driven prediction model for

Spark that predicts job performance with high accuracy by anticipating the execution time and memory usage of Spark applications.

## 3. METHODOLOGY

### 3.1. Application

Three common link prediction applications are chosen as the benchmark in our experiment, we developed three selected applications based on prevailing works that utilized the commonly used algorithms for scalable link prediction as discussed in Section 2. The three applications are Graph Clustering (GC), Overlapping Community Detection (OCD), and Redundant Graph Detection (RGD) use clustering-based, parallel label propagation-based, and path-based algorithms correspondingly. Therefore, the following is the pseudocode of the applications involved in our framework execution:

```
Algorithm 1: Graph Clustering, GC Application

Input: Edge lists file from HDFS pathFromHDFS, parallelism n, Spark context sc
Output: clusters of predicted link
Begin
val usersRDD, relationshipRDD = sc.textFile(pathFromHDFS)     // RDD creation
val graph = relationshipRDD.outerJoinVertices(usersRDD)       // generate new graph
pageRankGraph = graph.COMPUTE_PAGERANK(sc)                    // calculate PageRank value of each vertex
adamicAdarGraph = graph.COMPUTE_ADAMICADAR(sc)                // calculate similarity of each vertex
Repeat until convergence
clusteredGraph = adamicAdarGraph.COMPUTE_POWERITERATIONCLUSTER(k,sc)
clusteredGraph.foreach.print()       // list categorized cluster
```

```
Algorithm 2: Overlapping Community Detection, OCD Application

Input: Edge lists file from HDFS pathFromHDFS, parallelism n, Spark context sc
Output: overlapped communities
Begin
val usersRDD, relationshipRDD = sc.textFile(pathFromHDFS)     // RDD creation
val graph = relationshipRDD.outerJoinVertices(usersRDD)       // generate new graph
MaxIteration = m, Communities = n
graph = graph.AFFILIATE_COMMUNITIES(sc, n)     // community labelling on each vertex
Repeat until m<0
overlappedCommGraph = graph.DETECT_OVERLAPPINGCOMMUNITIES(m)   // detect overlapped comm
overlappedCommGraph.foreach.print()      // list detected communities
```

```
Algorithm 3: Redundant Graph Detection, RGD Application

Input: Edge lists file from HDFS pathFromHDFS, parallelism n, Spark context sc
Output: overlapped communities
Begin
val lines = sc.textFile(pathFromHDFS)     // RDD creation represent graoh records by lines
```

```
    Edges = lines.flatMapToPair()      // second RDD with JavaPairRDD API to form a connecting link
    triads = edges.groupByKey()     // combined both RDD to form triads (graph)
    trianglesWithDuplicates = triads.flatMap()
    uniqueTriangles = trianglesWithDuplicates.distinct()      // detect duplicated triads
    sout(uniqueTriangles)     // list out unique triangles
```

### 3.2. Self-configured framework

The Self-Configured Framework (SCF) is developed to be a part of the existing open-source Apache Spark framework as illustrated in Figure 1. For each submitted application to the framework, it will undergo an automated configuration beginning with collecting application details and cluster information, decision making for the best configuration, and completing with updating the best identified configuration. SCF used XGBoost classifier in the decision making process where we identify feature sets for XGBoost classifier, which classifies the feature sets into a suitable number of executors per node to be used by the application during execution. Based on the best classification results, the application is ready to be executed automatically through the update module.

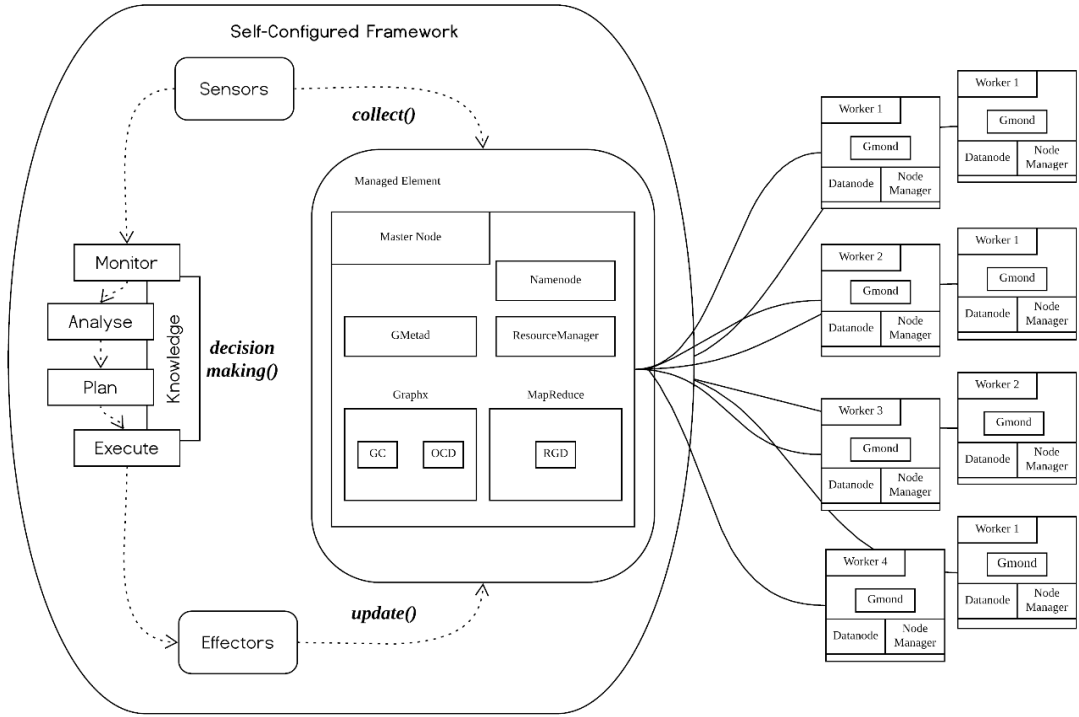

Figure 1: Full architecture of proposed Self-Configured Framework integrated with Apache Spark cluster

SCF consists of three modules and the pseudocode of the implemented modules are as follows:

| Algorithm 1: Collect, DecisionMaking, Update modules |
|---|
| **Module 1: Collect** <br> **Input:** P, path of the data file used in the analysis and application <br> **Output:** User configurations, application details, and cluster specification <br><br> 1. From the submitted link prediction application, compute P into size in megabytes, number of lines in the main class, and level of workload <br> 2. Acquire metadata of runtime cluster specification from the registered master server |

**Module 2: Decision Making**
**Input:** M: application metadata (*mm, mc, wn, wmn, wcn, ds, ac, mec*)
**Output:** New configuration of *driverCores*, *overheadDriverMemory*, *driverMemory*, *totalInstance*, *overheadMemoryPerExecutor*, *memoryPerExecutors*

1. Assign requested data from collect module to *M*
2. Dispatch *M* into http request API to access xgboost trained model
3. Update predicted value of *EPN* given *M* values in json form
4. if result = 0, make *EPN* = 1          // this is to handle small data cases
5. if result = 1, make *EPN* = 2
6. predefine upper bound values *MOC*, *EM*, *EC*, *ORC*, *ORM*, *PPC*
7. recalculate new configuration for application properties of *dc*, *odm*, *dm*, *ti*, *ompe*, *mpe* given *EPN* without trespassing upper bound

**Module 3: Update**
**Input:** New configuration properties
**Output:** Log updated configuration

1. Check cluster manager
2. If *cm* of submitted application is local, then
3. set smaller spark.driver.core to 2
4. If *cm* is standalone or YARN or Mesos or kubernetes, then
5. Set all recalculated configurations; spark.driver.core, spark.driver.memoryOverhead, spark.driver.memory, spark.executor.instances, spark.executor.memoryOverhead, spark.executor.memory, spark.executor.cores, spark.default.parallelism

### 3.3. XGBoost in Decision Making module

XGBoost is an implementation of gradient boosted decision tree algorithms, a sequential technique designed for speed and performance [12]. In Self-Configured Framework, we implement XGBoost trained model in Python as an API that returns Executor Per Node (EPN) value to DecisionMaking() module as shown in Figure 2. Technically, the input features for our XGBoost model include *masterMemory*, *masterCore*, *workerNode*, *workerMemoryNode*, *workerCoreNode*, *dataSize*, *applicationComplexity*, *memoryCapacity,* and the output or classified feature is predicted value of *executorPerNode*. Tuning parameters that we used are Learning Rate = *Range [0,1]*, Max_depth = *3*, N_estimators = *3*, Objective = *multi:softprob*. The training data used is 70 percent from a total of 16 million configuration records. Once the data set is divided into training and test sets, we build our model with randomly selected data points from the train set. Then we test the model using train set and achieve 100% accuracy. The trained model is then inserted into our storage space to be used by decision making module to request predicted EPN value.

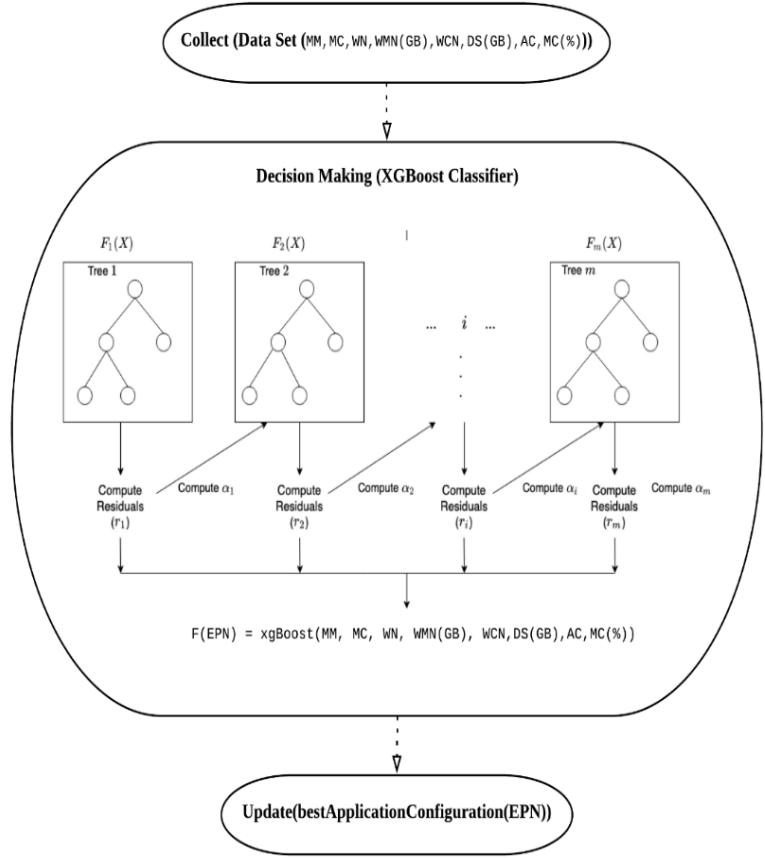

Figure 2: XGBoost model implementation as part of DecisionMaking() module in SCF

### 3.4. XGBoost vs other classifier

There are several classifications model we used namely; Decision Tree, Logistic Regression, K-Nearest Neighbors. Naïve Bayes, RF, and XGBoost. The best performance, accuracy of training in less time is the XGBoost model. As seen in Table 1, we list the accuracy, precision, recall F1 score, and training time taken for each classifier model. All of the models can achieve the correct classification of the executor per node.

Table 1: Accuracy and Training time comparison of XGBoost with other classifiers.

| Classifier | Accuracy | Precision | Recall | F1 Score | Training time(s) |
|---|---|---|---|---|---|
| **XGBoost** | 100% | 1.0 | 1.0 | 1.0 | 839 |
| **DT** | 100% | 1.0 | 1.0 | 1.0 | 133 |
| **LR** | 87% | 0.8 | 0.8 | 0.79` | 8682 |
| **KNN** | 100% | 1.0 | 1.0 | 1.0 | 3600 |
| **NB** | 86% | 0.81 | 0.73 | 0.74 | 6315 |
| **RF** | 100% | 1.0 | 1.0 | 1.0 | 2365 |

Three classifier models achieved 100% of accuracy, includes Random Forest, Decision Tree, and XGBoost. The reason for 100% accuracy is that the data used has no machine learning problem, we proposed to use a machine learning classifier to enhance the speed of self-configuring properties in SCF in comparison if we only apply a rule-based algorithm. The random forest has the longest training time that is more than half an hour. Decision Tree has the shortest training time, however, we picked XGBoost as our main classifier because XGBoost is the optimized version of decision tree and claimed to provide advanced learning techniques to yield superior results using fewer computing resources in the shortest amount of time [13].

Although it trained 10 minutes more than decision tree, is the best with multiple decision trees computation that requires more time than a single decision tree classifier. For the SCF dataset, both DT and XGBoost classifiers are reliable and suit tree-based algorithms.

## 4. SCF EVALUATION

The evaluation was conducted using Apache Spark with two different environments setup Azure Ubuntu 16.04 server for big cluster and Centos 7 ARM x86 server for a smaller cluster. Three types of cluster setup are 2, 4, and 8 node clusters configured in each environment as shown in Figure 3. Memory configurations range from 2GB to 32GB. Data used are 24 million edges and 73 thousand nodes in the form of edge lists. The applications considered in this evaluation are GC, OCD, and RGD written in Scala. The applications are all implementation of link prediction analysis in social networks detailed in Section 3.1.

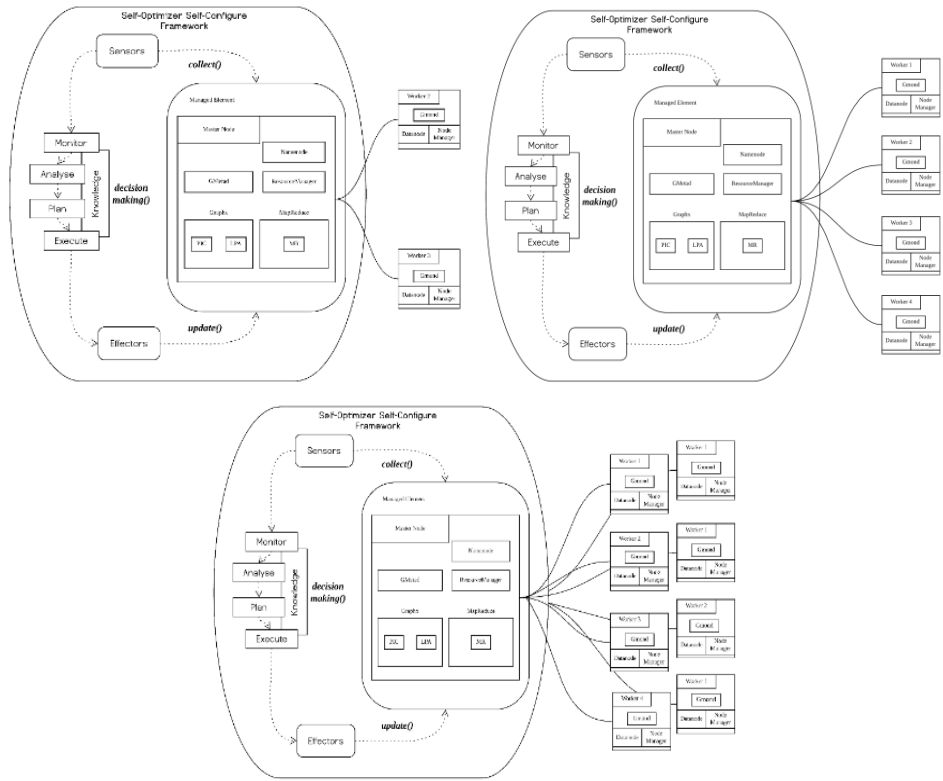

Figure 3: 2, 4, and 8 nodes of spark cluster setup for evaluation

We executed the evaluation with two different methodologies,

- The first evaluation was a comparison of the time performance and efficiency of the three applications in the default configuration and best configuration predicted by SCF.
- The second evaluation conducted a comparison of the resource utilization efficiency of each application.

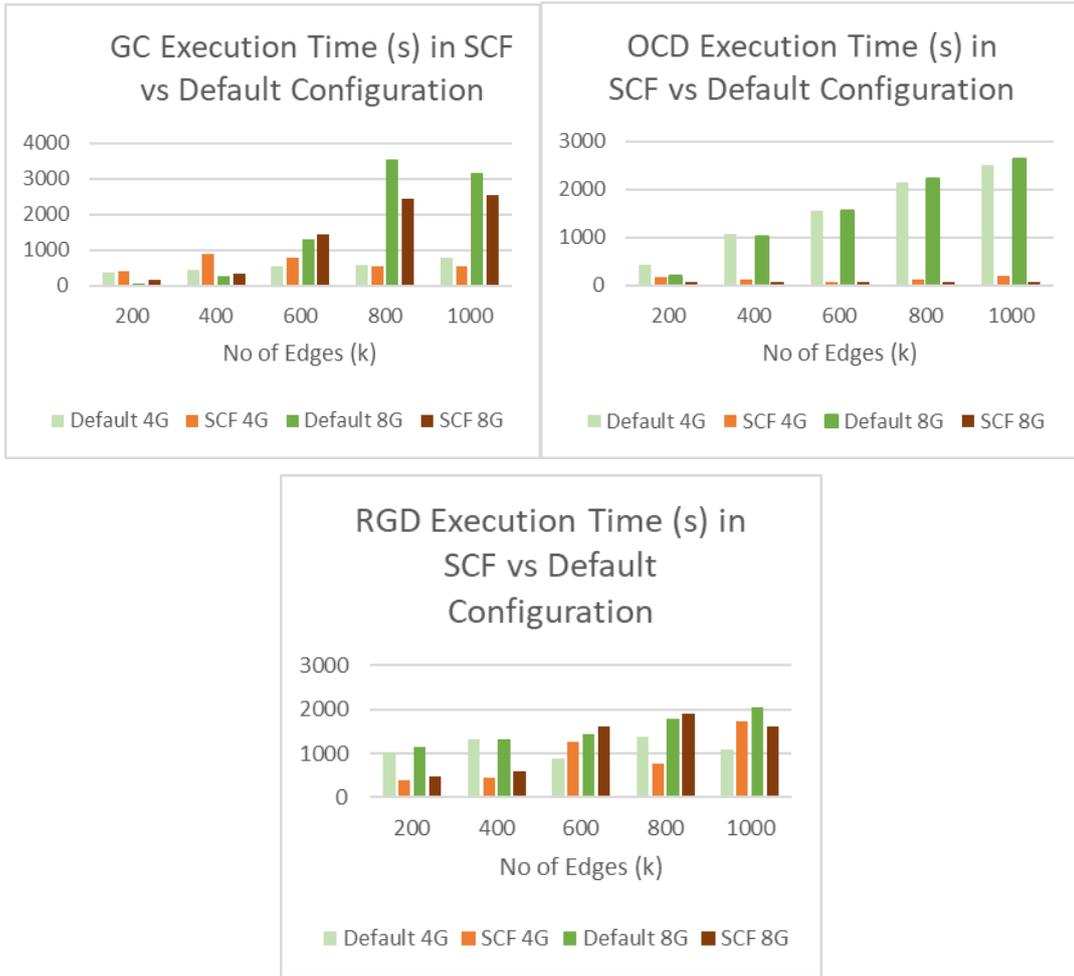

Figure 4: GC, OCD, and RGD execution time for each data used (K) in SCF vs Default Configuration

Figure 4 illustrated the time execution for each application executed in varying no of edges from 200k to 1000k with 4 different case studies which includes: Default configuration of 4GB and 8GB memory, SCF configuration of 4GB and 8GB memory. The plotted time shows that for OCD application, almost 40% lesser time taken to complete the execution when compare with the default configuration 4GB and 8GB memory case studies.

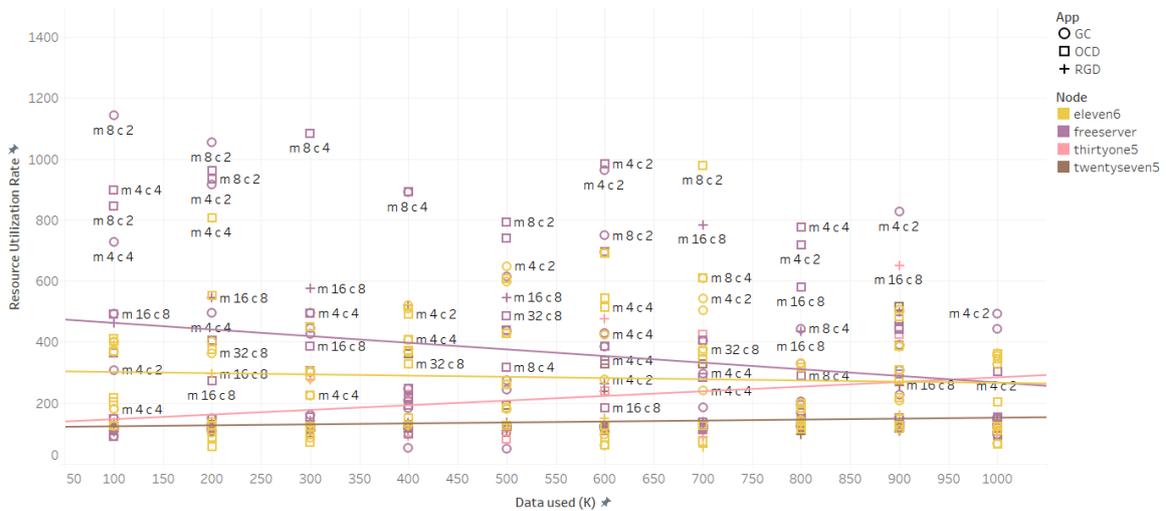

Figure 5: Resource utilization rate vs data used(K) for each application broken down by nodes

We also show how is the rate of all resource utilization in Figure 5. The resource utilization rate is the sum of all resource metrics monitored that are CPU utilization, memory usage in percent, and packet delivery ratio. We observe the resource utilization trend is almost similar for every resource used. This shows that SCF contributed to balancing resource utilization even in varying case studies.

## 5. CONCLUSIONS

This paper presented a Self-Configured Framework (SCF) integrated with Spark environment to refine the performance and inefficiency of link prediction in large-scale social networks, Twitter. The framework automatically configures the best configuration that suits to particular link prediction application given varying dataset size, workload, and cluster specification in Spark. To provide a general understanding of link prediction, this paper presented state-of-the-art of scalable link prediction in social networks. Further, this research proposed a newly generated dataset for spark clusters in order to invent a new combination set of features for predicting the best configuration of a certain case. Based on the set of features, SCF used XGBoost model to predict a suitable value of executor per node for a submitted link prediction application. The Self-Configured Framework is able to increase the execution time performance at almost 40% and balance the resource utilization when massive application is submitted to the framework for execution. However, SCF is developed as a framework-dependent because as of now it is integrated only with Apache Spark. Finally, this study can provide baseline information on the recent scalable link prediction applications in large-scale social networks the future of SCF would be to integrate with another framework like Hadoop, Storm for streaming analysis.

## ACKNOWLEDGEMENTS

We gratefully thank Dr. Zati Hakim Azizul Hasan for providing access to the Robotic Lab and ARM-based server. We are also fortunate to be allowed to conduct the experiment and development in paid servers, Azure Cloud sponsored by Assoc. Prof. Ts. Dr. Nor Badrul Anuar Bin Juma'at. Finally, we also thank Muntadher Saadoon for assistance in accessing and setting up the virtual environment on the ARM server.

## Authors

**Nur Nasuha Daud** received Bachelor degree in Computer Science and Information Technology (Software Engineering) from University of Malaya, Malaysia. Nasuha is currently pursuing a PhD program from the same university in the Department of Software Engineering, Faculty of Computer Science and Information Technology. Her Ph.D research is in Social network analysis with specific focus on Link Prediction.

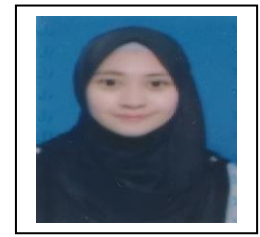

**Siti Hafizah Ab Hamid** received BS (Hons) in Computer Science from University of Technology, Malaysia, MS in Computer System Design from Manchester University, UK., and the PhD in Computer Science from University of Malaya, Malaysia. She is currently an Associate Professor with the Department of Software Engineering, Faculty of Computer Science & Information Technology, and University of Malaya, Malaysia. She has authored over 80 research articles in different fields, including mobile cloud computing, big data, software testing, software engineering, machine learning and IoT.

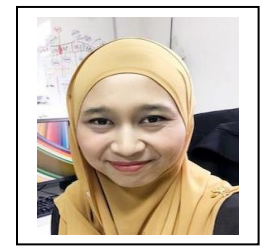

**Nor Badrul Anuar** received his Master of Computer Science from University of Malaya in 2003 and a PhD at the Centre for Information Security & Network Research, University of Plymouth, UK in 2012. He is currently an Associate Professor with the Faculty of Computer Science and Information Technology, University of Malaya. He has authored over 128 research articles and a number of conference papers locally and internationally.

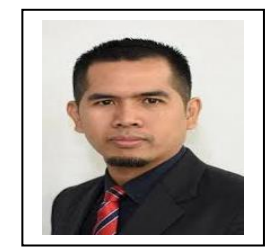